\documentclass{emulateapj}
\usepackage{psfig}
\usepackage{apjfonts}

\def\axp{1E~1048.1--5937}
\def\gsh{GSH~288.3--0.5--28}

\newcommand\HI{H\,{\sc i}}
\newcommand\HII{H\,{\sc ii}}
\def\kms{km~s$^{-1}$}

\shorttitle{ARE MAGNETARS FORMED FROM MASSIVE PROGENITORS?}
\shortauthors{GAENSLER ET AL}

\begin{document}
\title{A Stellar Wind Bubble Coincident with the Anomalous X-ray Pulsar
1E~1048.1--5937: Are Magnetars Formed From Massive Progenitors?}
\submitted{To appear in {\em The Astrophysical Journal (Letters)}}
\author{B. M. Gaensler,\altaffilmark{1} N. M. McClure-Griffiths,\altaffilmark{2}
M. S. Oey,\altaffilmark{3} M. Haverkorn,\altaffilmark{1}
J. M. Dickey\altaffilmark{4} and A.~J.~Green\altaffilmark{5}}
\altaffiltext{1}{Harvard-Smithsonian
Center for Astrophysics, 60 Garden Street MS-6, Cambridge, MA 02138;
bgaensler@cfa.harvard.edu}
\altaffiltext{2}{Australia Telescope National Facility, CSIRO, PO Box 76,
Epping, NSW 1710, Australia}
\altaffiltext{3}{Department of Astronomy, University of Michigan, 830
Dennison Building, Ann Arbor, MI 48103}
\altaffiltext{4}{Physics Department, University of Tasmania, GPO Box
252-21, Hobart, Tasmania 7001, Australia}
\altaffiltext{5}{{School of Physics, University of Sydney, NSW 2006, Australia}
}

\begin{abstract}

We present 21-cm \HI\ observations from the Southern Galactic Plane
Survey of the field around the anomalous X-ray pulsar \axp, a source
whose X-ray properties imply that it is a highly magnetized neutron star
(a ``magnetar'').  These data reveal an expanding hydrogen shell, \gsh,
centered on \axp, with a diameter of $35\times23$~pc (for a distance of
2.7~kpc) and an expansion velocity of $\approx7.5$~\kms. We interpret
\gsh\ as a wind bubble blown by a 30--40~M$_\odot$ star, but no such
central star can be readily identified.  We suggest that \gsh\ is the
wind bubble blown by the massive progenitor of \axp, and consequently
propose that magnetars originate from more massive progenitors than
do radio pulsars.  This may be evidence that the initial spin period
of a neutron star is correlated with the mass of its progenitor, and
implies that the magnetar birth rate is only a small fraction of that
for radio pulsars.

\end{abstract}

\keywords{
ISM: bubbles ---
ISM: individual (\gsh) ---
pulsars: individual (\axp) ---
radio lines: ISM ---
stars: winds, outflows ---
stars: neutron
}

\section{Introduction}
\label{sec_intro}

The last decade has revealed remarkable diversity in the young neutron
star population: radio pulsars, soft $\gamma$-ray repeaters (SGRS),
anomalous X-ray pulsars (AXPs) and central compact objects (CCOs) are all
now known to be potential compact remnants of core-collapse supernovae
\citep[e.g.,][]{kh02}. While the CCOs remain enigmatic, the AXPs and
SGRs are now both believed to be populations of ``magnetars'',
neutron stars whose persistent X-ray emission and occasional X- and
$\gamma$-ray bursts are powered by the energy associated with extreme
surface magnetic fields, $\ga10^{15}$~G \citep[see][for a review]{wt04}.

The low Galactic latitudes and the associations of some
of these sources with supernova remnants (SNRs) make it clear that
magnetars are young neutron stars \citep{gsgv01}. However, what
is not clear is why some neutron stars are ``normal'' radio pulsars,
while others are X- and $\gamma$-ray emitting magnetars. The physical
distinction cannot simply be the strength of the dipole magnetic field,
since there is now a small population of radio pulsars whose dipole
field strengths (as inferred from their spin down) are at magnetar-like
levels, but which have completely different X-ray characteristics
\citep{pkc00,km05}. Amongst the possibilities suggested to explain the
differences between these two populations are that the magnetic fields
of radio pulsars and magnetars differ in their geometry or orientation
\citep[e.g.,][]{zh00,kkm+03}, that magnetars originate from more massive
progenitors than do radio pulsars \citep{eml+04}, that high-field radio
pulsars are quiescent magnetars \citep{km05}, or that radio pulsars and
magnetars form an evolutionary sequence \citep{lyn04,hh05}.

The environments of magnetars provide unique constraints on their
origin and evolution. Here we present \HI\ observations of the
interstellar medium (ISM) surrounding the AXP \axp\ \citep[][\& references
therein]{gk04}, which show that this source is possibly associated with a
large expanding \HI\ shell.  This association strengthens the possibility
that magnetars are formed from more massive progenitor stars than are
radio pulsars.

\section{Observations}
\label{sec_obs}

The \HI\ data shown here were taken from the Southern Galactic
Plane Survey \citep[SGPS;][]{mdg+05}, which has mapped \HI\ and
1.4-GHz continuum emission in the range $253^\circ \le l \le 358^\circ$,
$|b| \le 1\fdg5$, at a resolution of $\approx2'$. The survey combines
interferometric data from the Australia Telescope Compact Array with
single-dish data from the Parkes 64-m radio telescope, giving
sensitivity to a wide range of spatial scales.  The final spectral line
data-set has a velocity resolution of $\sim0.8$~\kms\ and a sensitivity
of $\sim2$~K.

\begin{figure}
\centerline{\psfig{file=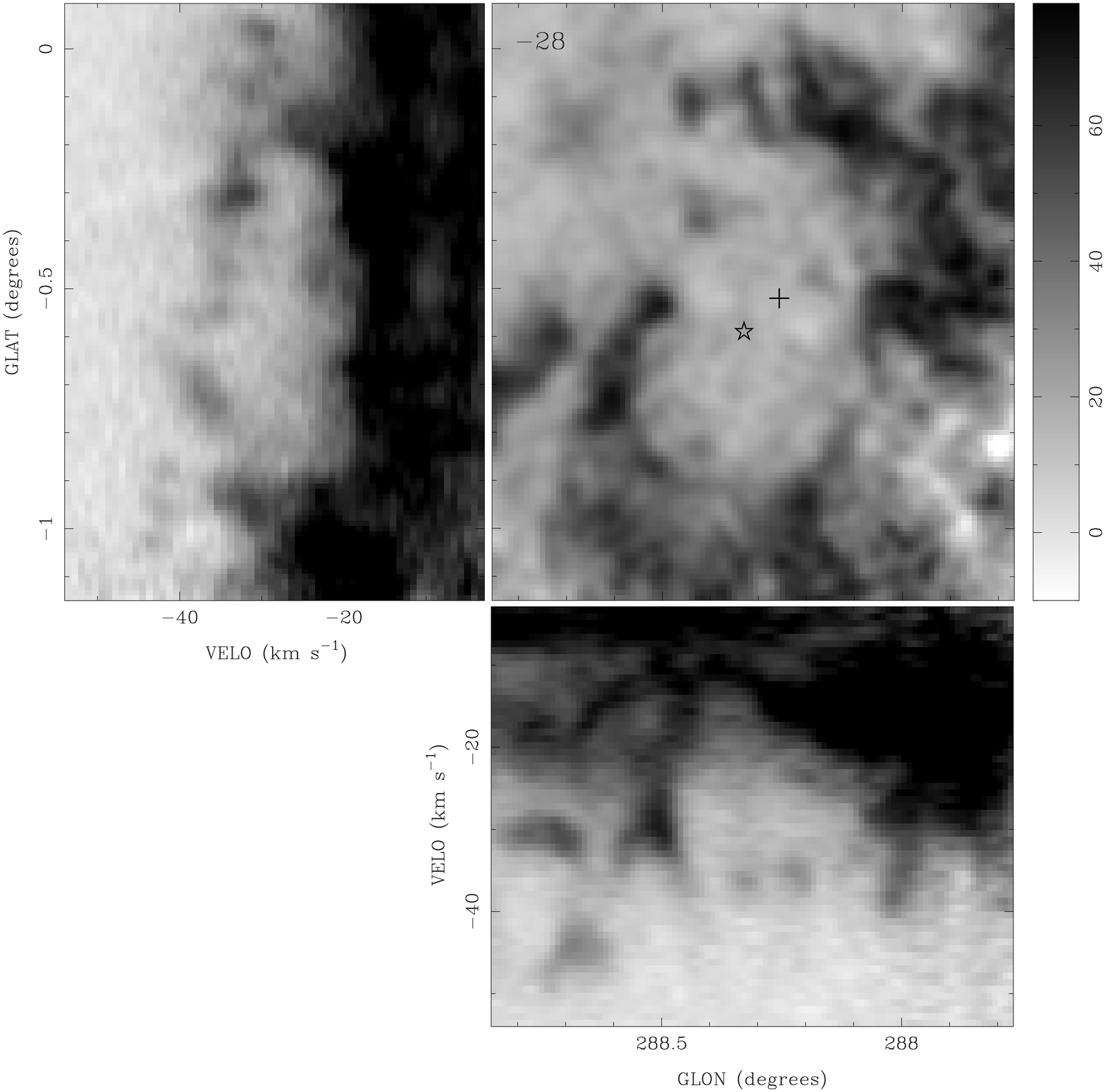,width=\textwidth}}
\caption{\HI\ in the field of the AXP~\axp, from SGPS data.  The three
panels plot the surface brightness of \HI\ as a function of Galactic
longitude, Galactic latitude and LSR velocity. Each greyscale ranges
between --10 and +78~Jy~beam$^{-1}$, as shown by the scale bar on the
upper right. The spatial resolution of the data are $142''\times125''$,
while the velocity resolution is 0.82~\kms. The upper right panel shows
the image plane at an LSR velocity of --28~\kms; the positions of \axp\
and of the star IX~Car are marked by a cross and a star, respectively.
The left
panel shows a $b-v$ diagram at $l = 288\fdg25$; the lower
panel shows a $l-v$ diagram at $b = -0\fdg52$.}
\label{fig_bubble}
\end{figure}

The SGPS reveals a striking cavity in \HI, almost centered on the
position of \axp\ (namely $l = 288\fdg3, b=-0\fdg5$), as shown in
Figure~\ref{fig_bubble}.  The shell has a central velocity relative to
the Local Standard of Rest (LSR) $V = -28$~\kms; we thus designate this
object \gsh. At this central velocity, the major and minor axes of \gsh\
are $45'$ and $29'$, respectively.  The images of $l$ versus $V$ and $b$
versus $V$, also shown in Figure~\ref{fig_bubble}, demonstrate that \gsh\
is expanding, at a velocity $V_{exp} \approx 7.5$~\kms.  \gsh\ is seen
in multiple velocity planes, shows a high contrast between its walls
and interior, and changes in angular extent as a function of velocity.
It thus meets all the standard criteria for \HI\ shells in the ISM
\citep[see][]{mdgg02}.

Standard rotation curves \citep[e.g.,][]{bb93} forbid systemic motions
having the central velocity observed for \gsh, indicating significant
deviations from circular rotation in this region.  However, the shell's
central velocity matches the  {\em observed}\ terminal velocity in this
direction, which allows us to estimate a distance of 2.7~kpc from geometry
alone.  We assume that random cloud motions on the order of 6~\kms\
contribute to an error in the distance estimate on the order of 40\%, or
1~kpc.  Incorporating both this distance uncertainty and the non-circular
shape of the shell, we adopt a radius $R = 14\pm7$~pc.  By integrating
along a number of lines of sight through the shell walls and then
subtracting a mean background, we find an average hydrogen column density
through the shell walls of $(2.5\pm0.6)\times10^{20}$~cm$^{-2}$. If this
material uniformly filled the region into which the shell has expanded,
the implied ambient number density is $n_0\approx17\pm9$~cm$^{-3}$.

\section{Interpretation}
\label{sec_int}

\subsection{The Nature of \gsh}
\label{sec_int1}

Shells like \gsh\ are common, and result from the effect of massive
stars on the ISM.  Specifically, \gsh\ could be an \HII\ region, a SNR,
or a stellar wind bubble.  Below we briefly consider these
possibilities

The photoionizing flux needed to maintain an \HII\ region is $N_*
= 4 \pi R^3 n_0^2 \alpha$/3, where $\alpha \approx
3\times10^{-13}$~cm$^3$~s$^{-1}$ is the recombination coefficient
for hydrogen \citep{ost89}.  For the observed values of $R$ and
$n_0$, we require $N_* \sim (3\pm2) \times 10^{49}$~s$^{-1}$.  This
is consistent with the ionizing fluxes of the earliest-type stars,
but such a source would also have a radio flux of $>100$~Jy at
decimeter wavelengths.  A radio continuum image of this region shows
no such emission associated with \gsh\ \citep{gcly99}.

For the observed values of $n_0$, $R$ and $V_{exp}$, a SNR in the
Sedov or radiative phases requires an initial explosion energy of
$\sim5 \times10^{49}$ or $\sim10^{50}$~ergs, respectively. These
estimates are both well below the typical supernova explosion energy
of $10^{51}$~ergs, making this interpretation unlikely. There is
also no evidence in deep archival observations for any radio or
X-ray SNR associated with \axp\ \citep[see][]{gsgv01}.

The remaining possibility is that \gsh\ is a wind-driven bubble.
Using the wind bubble solution of \cite{wmc+77}, during
the energy-conserving phase of expansion we expect that
$R = 0.76 (L_w/ \rho_0)^{1/5} t^{3/5}$,
where $t$ is the age of the bubble, $L_w$ is the wind luminosity
of the central star, and $\rho_0$ is the density of the
ambient medium.  Since $V_{exp} = dR/dt$, we find 
$L_w \approx (4\pm2)\times10^{35}$~ergs~s$^{-1}$ and $t=1.1\pm0.5$~Myr.
The luminosity of the wind is a strong function of mass,
allowing a reasonable typing of the associated central object (assuming
only one star powers this bubble).  In this case, an appropriate
central source is an O6V star, with a mass-loss rate $\dot{M} \approx 2
\times 10^{-7}$~M$_\odot$~yr$^{-1}$ \citep{dnv88}, a wind velocity 
$\approx 2500$~\kms\ \citep{pbh90}, and a zero-age main sequence (ZAMS)
mass of 30--40~M$_\odot$ \citep[e.g.,][]{mpv02,ol03}.  

\subsection{\gsh\ and Carina~OB1}
\label{sec_car}

We have shown that \gsh\ is most likely powered by the wind of a
massive star at a distance of $2.7\pm1$~kpc. But can we identify
the central star responsible for this bubble? There are many massive
stars in this direction and at this distance, most of which are
part of the association Carina~OB1 at a distance of $\sim2.5$~kpc
\citep[e.g.,][]{hum78}. However, we now argue that \gsh\ is
unassociated with, and probably more distant than, the stars in
Car~OB1.

First, the collective winds of the massive stars in Car~OB1 correspond
to a total luminosity of $\sim8\times10^{37}$~ergs~s$^{-1}$
\citep{abb82b}. Combined with the powerful winds from stars in the
nearby clusters Tr~14 and Tr~16, this should power a much larger
expanding superbubble. Indeed such a larger shell, 100--200 pc
across, has been identified in ionized, neutral and molecular gas
\citep{chty81,ra98}.  Any individual star in this vicinity will not
sweep up a neutral shell of its own but will contribute to the
overall ionized supershell.  Since the properties of \gsh\ are
consistent with a single star sweeping up neutral gas, it seems
unlikely that this shell is in the vicinity of Car~OB1.

Second,  the visual extinction toward the stars in Car~OB1 is low,
$A_V \la 3$ \citep{hum78,hlv92}. At this extinction and distance,
a massive star should be clearly detected at the center of \gsh.
For example, the O6V star considered above would have a magnitude
$V \la 9$.  We have performed an exhaustive search for
such sources \cite[e.g.,][]{hum78,fo81,ree98,kal03}, from which the
only catalogued massive star projected near the center of \gsh\ is
IX~Car (HD~94096; see Fig.~\ref{fig_bubble}), an M2Iab star with a
mass of $\sim20$~M$_\odot$.  The distance estimated to IX~Car of
$\sim1.6$~kpc and its LSR radial velocity of  --17~\kms\ \citep{hum78}
suggestthat it is a foreground object. Other massive stars in the
vicinity, such as HD~93843, HD~94230, HD~305599 and LS~1976, all
lie on the very edge or outside of \gsh\ in projection.

Finally, the photoionization from stars in Car~OB1, most notably
the O5III star HD~93843, should fully ionize the shell. If we assume
that the shell thickness is 20\% of its radius, the edges of the
shell should show H$\alpha$ emission at a surface brightness of
thousands of rayleighs. We have examined sensitive images of this
region \citep[e.g.,][]{bbw98,gmrv01}; these show no H$\alpha$
emission associated with \gsh\ down to much lower surface brightness
limits. We therefore believe that \gsh\ is not exposed to the high
ionizing fluxes associated with Car~OB1, but likely has a higher
extinction than and is behind these coincident stars.

An important point is that the high extinction that we have invoked
only need imply a slightly larger distance for the \HI\ shell than for
Car~OB1, since extinction in this direction has little correspondence with
distance.  This is clear from consideration of the $\sim30$ Wolf-Rayet
(WR) stars within $3^\circ$ of \gsh\ \citep{vdh01}. For stars from this
sample with distances in the range 1.7--3.7~kpc as appropriate
for \gsh, $A_V$ can be as high as 6--7.

\subsection{The Central Source Associated with \gsh}

If \gsh\ is not associated with any of the known bright stars in
this direction, what is powering it? If the extinction to \gsh\ is
significantly higher than to the stars in Car~OB1, its associated
star might not be obvious.  We first consider K or M supergiants,
which should be prominent in the near-infrared --- we estimate a
1.3-$\mu$m magnitude $J \la 6.5$ for a red supergiant at the distance
to \gsh. However, using the 2MASS point source catalog \citep{csv+03},
we find no candidates for such a source inside \gsh, other than IX~Car
(see \S\ref{sec_car}).  We searched for faint OB stars by considering
the Tycho-2 catalog \citep{hfm+00}, which is 90\% complete down to a $V$
magnitude of 11.5. We assume that any massive star with $A_V \la 4$ in
this well-studied region would have been classified in earlier efforts,
and so only consider stars with $A_V \ga 4$. For standard reddening,
this implies a color excess $E_{B-V} \ga 1.3$ and hence for OB stars $B-V
\ge 1$.  Just nine such stars from the Tycho-2 catalog lie within the
perimeter of \gsh, of which three are K or M stars with low extinction,
four have high proper motions which indicate that they are foreground
objects, one is a foreground B1V star (LS~1956) which is heavily reddened,
and one has near infrared magnitudes from 2MASS which are inconsistent
with it being a massive star at the distance to \gsh.

It is certainly possible that  stars too faint to be in the Tycho-2
catalog could also be candidates.  While we thus accept that we cannot
make a definitive identification using available data, there is one
remaining source, namely the AXP~\axp, which we argue below represents
a viable association with \gsh.  If \axp\ generates a wind powered by
its spin-down, as do many radio pulsars, then its spin parameters imply
only $L_w \sim4\times10^{33}$~ergs~s$^{-1}$ \citep{gk04}.  This is far too
low to be responsible for the surrounding shell.  However, an intriguing
alternative is that \gsh\ was blown by the star whose collapse
formed \axp.  As for associations between pulsars and SNRs, the likelihood
of this association should be judged by geometric correspondence,
agreement in distance and age, and evidence for physical interaction.

First considering geometry, it is clear from Figure~\ref{fig_bubble}
that \axp\ is close to the
center of \gsh\ in projection. Isolated neutron stars generally have
high space velocities, $\ga 300$~\kms \citep{acc02},\footnote{We note
that no velocity measurements yet exist specifically for magnetars
\citep{gsgv01}.} so that
any neutron star older than $\sim100$~kyr will
have moved far from its progenitor's wind bubble.  
Thus the association is only viable if
we can argue independently that \axp\ is extremely young. Indeed there
is good evidence  that this is the case, since associations with SNRs
for other AXPs argue that these sources all have ages
$\la10$~kyr \citep{ggv99,gsgv01}.  In this case the AXP should
be centrally located in any progenitor wind bubble, as observed.
We note that the ``characteristic age'' of \axp,
$P/2\dot{P} \approx 4$~kyr \citep{gk04},
is indeed small (where $P$ is the star's spin period).\footnote{\axp\
experiences significant torque variations; we have adopted the
long-term average for $\dot{P}$.}
However, we caution that this and other
magnetars do not spin down like normal radio pulsars \citep{tdw+00,gk04},
so this age estimate does not provide independent evidence for youth.

Second, we consider distance estimates. The systemic velocity of
\gsh\ puts it at $2.7\pm1$~kpc. No distance estimates exist for \axp,
except for the fact that the absorbing column of hydrogen inferred
from its X-ray spectrum, $N_H \approx 1.1\times10^{22}$~cm$^{-2}$
\citep[e.g.,][]{tgsm02}, is much higher than that for the Carina complex
(which includes Car~OB1) at a distance of $\approx2.5$~kpc. It has
thus been argued that 2.5~kpc is a firm lower limit on the distance to
this source \citep{opk01}. However, the high column to \axp\ is still
compatible with the distance to \gsh. For standard gas to dust ratios,
$N_H \approx 1.1\times10^{22}$~cm$^{-2}$ implies a visual extinction $A_V
\approx 5.8$ \citep{wc02}.  The WR catalogue of \cite{vdh01} demonstrates
that this level of extinction is consistent with distances in the range
$\sim2-12$~kpc in this region, so is easily reconciled with the distance
to \gsh.

Third, we need to determine if the age of \gsh\ is consistent with
that expected for the progenitor of \axp.  A star with a ZAMS mass of
$\sim30-40$~M$_\odot$ and solar metallicity lives for $\sim5-6$~Myr
\citep[e.g.,][]{mms+94}.  Since the age of the AXP is negligible in
comparison, this should also be the age of the surrounding 
bubble.  However, in \S\ref{sec_int1}, we esimated an age for the shell
of $1.1\pm0.5$~Myr. There is clearly a large discrepancy between these two
estimates. However, this same age problem has been observed for many other
\HI\ shells around individual massive stars \citep[e.g.,][]{gs99,ch00b}.
There is no simple resolution to this discrepancy, although in some cases
it may be explained by a ``blow out'' into a lower density environment
\citep{os98}. We conclude that any discrepancy between the ages estimated
for \gsh\ and for the progenitor of \axp\ is not a strong argument
against their association.

Finally, we consider direct physical evidence that \axp\ and \gsh\
are associated.  Young neutron stars should be embedded in young
SNRs, but for \axp\ no such SNR is observed \citep{gsgv01}.  A
simple explanation is that the associated SNR is expanding into a
low density bubble, so that it does not yet produce observable
emission \citep[e.g.,][]{cd89}.  Thus from the absence of a SNR,
it is reasonable to expect that \axp\ should show evidence for a
surrounding cavity:  \gsh\ clearly fulfills this prediction. For a
SNR shock velocity of 5000~\kms, the blast wave should impact the
shell walls $\sim3$~kyr after core collapse.  The lack of any radio
or X-ray emission from this event requires the neutron star to be
younger than this, consistent with the small ages expected for AXPs.

To summarize, a very young neutron star with a massive progenitor should
be centrally located in an expanding \HI\ shell, with no evidence for a
surrounding SNR. This is exactly what we observe for \gsh\ and \axp. Both
sources are consistent with being at a distance of $\sim3$~kpc, behind
$\sim6$ magnitudes of optical extinction.

\section{Implication for Magnetars}
\label{sec_mag}

As explained in \S\ref{sec_intro}, a fundamentally unresolved issue
in the study of compact objects is why some neutron stars
are ordinary radio pulsars, while others are magnetars.  Since the initial
mass function (IMF) sharply declines with increasing ZAMS mass, most
neutron star progenitors will have masses near the minimum mass for
core collapse, i.e., 8--9~M$_\odot$.  We have presented evidence
here that the progenitor of AXP~\axp\ was considerably more massive than
this.  For some SGRs, possible associations with massive star clusters
similarly argue for high-mass progenitors \citep[e.g.,][]{khg+04,eml+04}.

We thus propose that the difference between normal pulsars and
magnetars is the progenitor mass. We note that massive ($\ga
25$~M$_\odot$) stars do not always form black holes: for solar
metallicity, mass loss causes single stars heavier than
$\sim60$~M$_\odot$ to form neutron stars \citep{hfw03}; a progenitor
of lower mass ($25-40$~M$_\odot$) whose binary companion strips
its outer envelope before core collapse can also form a neutron
star \citep{fk01}.

Why should massive stars form magnetars? \cite{dt92a} and \cite{td93a}
argue that magnetars result from rapidly rotating ($P \sim 1$~ms)
proto-neutron stars, in which an efficient large-scale dynamo operates
in the first few seconds after birth, generating a super-strong magnetic
field with significant multipolar components.  \cite{hws04} have recently
carried out a series of calculations of differentially rotating magnetized
supernova progenitors.  They find that magnetic torques are especially
effective at transferring angular momentum away from the stellar core
during the red supergiant and helium burning phases of evolution.
For progenitors with masses $\sim10-15$~M$_\odot$, this results in
neutron stars with initial periods $\sim10-15$~ms, too slow to generate
magnetar-like fields.  However for more massive stars, the reduced
time interval between hydrogen depletion and the supernova results in
limited braking of the core, producing much more rapidly spinning neutron
stars. For example, a 35~M$_\odot$ progenitor results in a neutron star
of initial period 3~ms, in the range needed to give birth to a magnetar.

Within this scenario, we can immediately make a prediction as to the
relative birth rates of radio pulsars versus magnetars.  For argument's
sake, we suppose that the mass cut between normal pulsars and magnetars
is at a ZAMS mass of 25~M$_\odot$. Then for an IMF of slope $\alpha =
2.35$ for massive stars ($dN/dM \propto M^{-\alpha}$; \citeauthor{mas03}
\citeyear{mas03}), only $\sim20$\% of progenitors can potentially form
magnetars. Since $\sim50$\% of progenitors heavier than 25~M$_\odot$
will form black holes \citep{hfw03}, we infer that the magnetar birth
rate is $\sim10$\% of that of radio pulsars.  The birth rate estimated
from observations of the existing populations of AXPs and SGRs is already
comparable to this \citep{kfm+04,ggv99}. If magnetars indeed form from
massive progenitors, we anticipate that there is no substantial population
of such sources yet to be identified \cite[cf.,][]{ims+04}.

\begin{acknowledgements}
We thank Cristina Cappa, Chris Fryer, Andrew Melatos and Chris Thompson
for useful discussions, and Paul Price for supplying H$\alpha$ data.
The Australia Telescope is funded by the Commonwealth of Australia for
operation as a National Facility managed by CSIRO.  This paper has
used the CDS SIMBAD and VizieR Catalogue Services,  and 
NASA's ADS and SkyView facilities.  B.M.G. is supported by NASA through
LTSA grant NAG5-13032.
\end{acknowledgements}


\end{document}